\documentclass[twocolumn,aps,showpacs,amsmath,amssymb,preprintnumbers,epsf]{revtex4}
\usepackage{graphicx}
\usepackage{dcolumn}
\usepackage{color}
\usepackage{bm}
\def\eqn{\end{equation}\noindent}
\def\eqnr{\end{eqnarray}\noindent}
\def\beqr{\begin{eqnarray}}
\def\beq{\begin{equation}}

\def\beq{\begin{equation}}

\def\beq{\begin{equation}}
\def\eqn{\end{equation}}
\begin {document}

\title{\textbf{Measurement of phase synchrony of coupled segmentation clocks}}
\author{Md. Jahoor Alam$^1$, Latika Bhayana$^1$, Gurumayum Reenaroy Devi$^1$, Heisnam Dinachandra Singh$^1$, R.K. Brojen Singh$^1$, B. Indrajit Sharma$^2$.}
\affiliation{$^1$Centre for Interdisciplinary Research in Basic Sciences, Jamia Millia Islamia, New Delhi 110025, India\\$^2$Department of Physics, Assam University, Silchar 788 011, Assam, India\\
}
\date{\today}

\begin{abstract}

{\begin{center}\bf ABSTRACT\end{center}}
{The temporal behaviour of segmentation clock oscillations show phase synchrony via mean field like coupling of delta protein restricting to nearest neighbours only, in a configuration of cells arranged in a regular three dimensional array. We found the increase of amplitudes of oscillating dynamical variables of the cells as the activation rate of delta-notch signaling is increased, however, the frequencies of oscillations are decreased correspondingly. Our results show the phase transition from desynchronized to synchronized behaviour by identifying three regimes, namely, desynchronized, transition and synchronized regimes supported by various qualitative and quantitative measurements. 
}
\\\\
\emph KEYWORDS
{: Delta-notch coupling, phase synchrony, Hilbert transform, Phase locking value, mean field coupling.}
\end{abstract}
\maketitle

\section{Introduction}

The origin of oscillatory behaviour of segmentation clock genes is believed to be due to negative feedback genetic regulation of their own products \cite{giu,hir,hol}. For example, in vertebrate developement, periodic formation of somites during presomitic mesoderm is considered to be responsible to exhibit oscillatory expression by these genes during sometogenesis and their period of oscillation is very close to the period of segmentation\cite{bes,hol1,jou,oat,pal}. It has been reported that during this signal processing neighbouring cells are in contact and these cells communicate among the neighbouring cells showing synchronization in the dynamical variables \cite{mar,mas}.

There have been various theoretical studies regarding the dynamical stability, periodicity and synchronization of the various oscillating variables of segmentation clock \cite{coo,bak,cin,golb,uriu,hor,lew}. Since synchronization and oscillation in the clock gene regulations are believed to be responsible for somite boundary, these properties are considered to be essential in segmentation process \cite{ozb}. If a group of such cells are considered, signal processing among the neighbouring cells are done via delta-notch signaling process to achieve synchronization among them \cite{hor,ozb,uri3}. However, since the delta proteins cannot diffuse from one cell to another, the incorporation of delta-notch coupling in the theoretical models so far developed is still questionable. One attempt in this direction without considering time delay was to introduce a coupling term in the model associated with activation rate of delta-notch signaling which enable to control the fluctuation in delta protein concentration in its neibouring cells \cite{uri1,uri2}. It has been reported that the cells in this coupling mechanism exhibit phase synchrony in one and two dimensional array of such cells \cite{uri1,uri2,uri3}.

In this work we focuss on some issues namely, measure of phase synchrony among the cells and generalization of coupling mechanism into some more real situation. In the section II, we study the model of segmentation clock gene expression due to Uriu et.al. (2009). We generalized the model in three dimensional situation with coupling mechanism. Then we explain various methods to identify phase synchrony among the coupled cells. The results and discussions of our simulation results are presented in section III. We draw some conclusions based on our results in section IV.
\begin{figure}
\label{fig1}
\begin{center}
\includegraphics[height=110 pt]{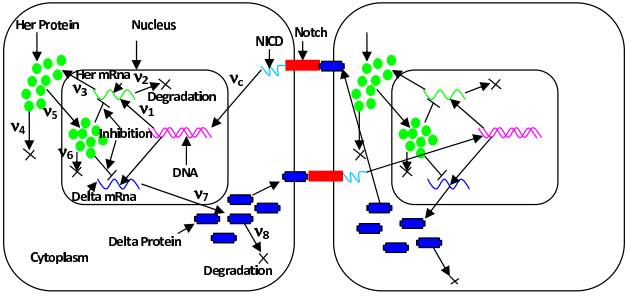}
\caption{The schematic diagram of reaction network of cell to cell interaction via delta-notch coupling mechanism.}
\end{center}
\end{figure}
\begin{figure}
\label{fig2}
\begin{center}
\includegraphics[height=130 pt]{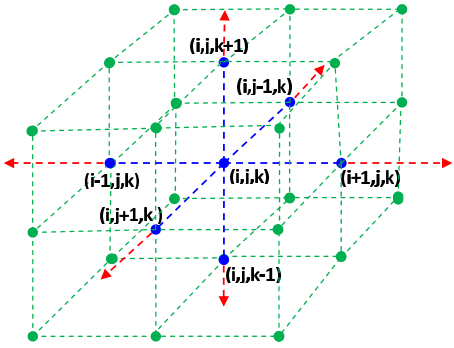}
\caption{The three dimensional array of cells showing nearest neighbour interaction.}
\end{center}
\end{figure}

\section{Materials and methods}

In vertebrate sometogenesis, the expression of ${\it her}$ gene in single cell model with and without time delay is essential in order to generate oscillatory behavior of the segmentation clock genes \cite{hor,bar, tie, uri1}. Here, we consider a model in
which simple kinetics of mRNA, protein in cytoplasm and protein in nucleus are considered without taking into account any time delay \cite{uri2}. The detail molecular reaction pathways are shown in Fig. 1.
\begin{figure}
\label{fig3}
\begin{center}
\includegraphics[height=170 pt]{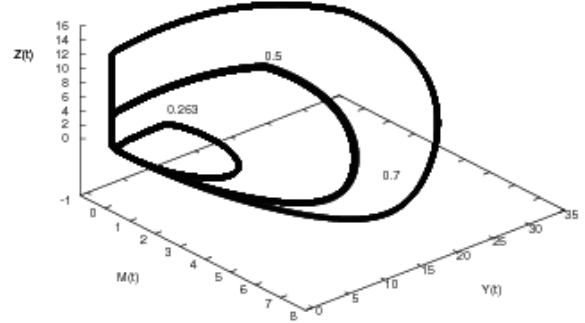}
\caption{The three dimensional plot of $M(t)$, $Y(t)$ and $Z(t)$ for $\nu_c=0.263,0.5$ and 0.7. The other values of the parameters used in the simulation are; $\nu_1=0.0135$, $\nu_2=0.6$, $\nu_3=0.575$, $\nu_4=0.851$, $\nu_5=0.021$, $\nu_6=0.162$, $\nu_7=2.465$, $\nu_8=9.583$, $k_1=0.157$, $k_2=0.104$, $k_4=0.142$, $k_6=0.13$, $k_7=0.49$, $k_8=9.72$,$n=h=2$ \cite{uri3}.}
\end{center}
\end{figure}

In this model \cite{uri2}, if $M$, $Y$, $Z$ and $W$ indicate concentrations of $Her$ mRNA, $Her$ protein in cytoplasm, $Her$ protein in nucleus and Delta protein respectively in a single cell, then the time evolution of these variables without considering time delay in describing the kinetics in the cell are given by,
\begin{eqnarray}
\frac{dM}{dt}&=&(\nu_1+\nu_cW)\Gamma_1-\nu_2\Gamma_2\\
\frac{dY}{dt}&=&\nu_3M-\nu_4\Gamma_3-\nu_5Y\\
\frac{dZ}{dt}&=&\nu_5Y-\nu_6\Gamma_4\\
\frac{dW}{dt}&=&\nu_7\Gamma_5-\nu_8\Gamma_6
\end{eqnarray}
where, $\Gamma_1=\frac{k_1^n}{k_1^n+M^n}$, $\Gamma_2=\frac{M}{k_2+M}$, $\Gamma_3=\frac{Y}{k_4+Y}$, $\Gamma_4=\frac{Z}{k_6+Z}$, $\Gamma_5=\frac{k_7^h}{k_7^h+Z^h}$ and $\Gamma_6=\frac{W}{k_8+W}$.
The parameters in the differential equations are given as follows: $\nu_1$ is the basal transcription rate, $\nu_c$ indicates the activation rate due to Delta-Notch
signaling, $\nu_2$ is maximum degradation rate of $Her$ mRNA, $\nu_3$ refers to
translation rate of Her protein, $\nu_4$ is the maximum degradation rate of $Her$
protein in cytoplasm, $\nu_5$ indicates transportation rate of $Her$ protein,
$\nu_6$ is maximum degradation rate of $Her$ protein in nucleus, $\nu_7$
is the sysnthesis rate of Delta protein, $\nu_8$ is the maximum degradation rate
of Delta protein, $k_1$ is threshold constant of supression of $Her$ mRNA
transcripted by $Her$ protein, $k_2$ is Michaelis constant for $Her$ mRNA
degradation, $k_4$ is Michaelis constant for $Her$ protein degradation in
cytoplasm, $k_6$ is Michaelis constant for $Her$ protein degradation in nucleus,
$k_7$ is threshold constant for the suppression of Delta protein synthesis by
$Her$ protein, $k_8$ is Michaelis constant for Delta protein degradation
in nucleus.

A group of such cells do communicate via delta-notch coupling mechanism when the cells are very close to each other or almost in contact each other. Since Delta-proteins are confined only on the cell membrane and cannot diffuse from one cell to another, signal is considered to pass from one cell to another by activating the delta-notch signaling pathway when the cells are in contact or very close. We consider $(N\times N\times N)$ cells which are arranged in a regular three dimensional array in this type of communication, assuming the sizes of the cells are the same as shown in Fig. 2. In this situation, any cell is allowed to couple with six nearest neighbour cells. Since the coupling variable $W$ is common to $(N\times N\times N)$ identical cells, this species can be an effective means of coupling each cell to every other via $\frac{1}{L}\sum_{i=1}^{L}W_i$, where, $L=6$ and summation is restricted to nearest neighbours only for every cells. The common species provides a "localized" mean field like coupling  whereby the cells communicate, and the remaining variables synchronize.
\begin{figure}
\label{fig4}
\begin{center}
\includegraphics[height=175pt]{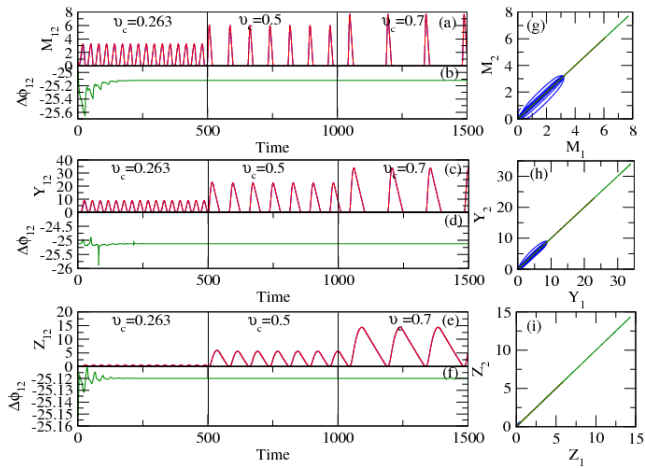}
\caption{Plots of the variables $M(t)$, $Y(t)$, $Z(t)$ as a function of time. The same parameters are used as in Fig. 3.}
\end{center}
\end{figure}
\begin{figure*}
\label{fig5}
\begin{center}
\includegraphics[height=370 pt]{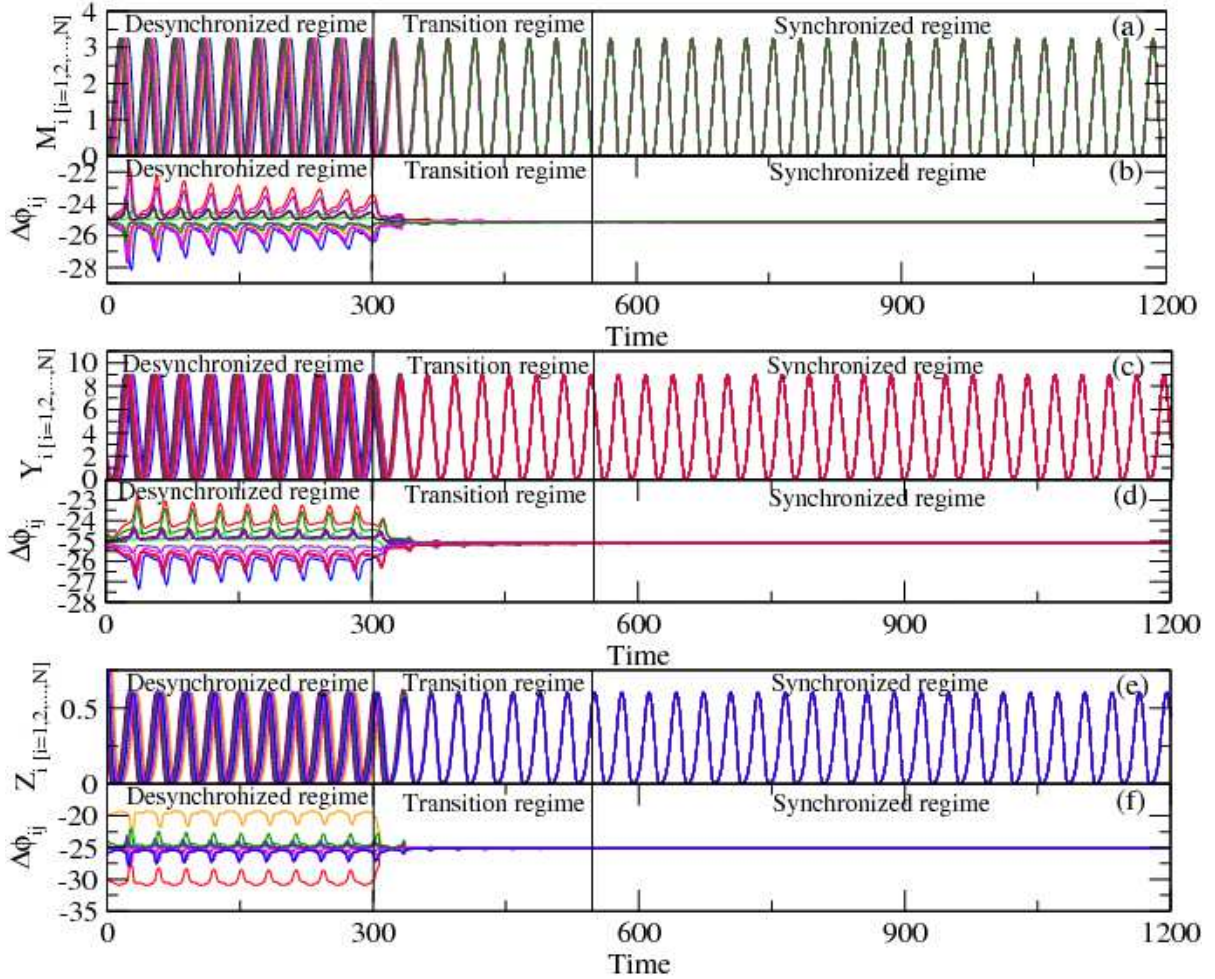}
\caption{Plots of time evolution of $M(t)$, $Y(t)$, $Z(t)$ of 10 oscillators out of $(10\times 10\times 10)$ three dimensional array of oscillators are shown in (a), (c) and (e). The corresponding plots of $\Delta\phi_{ij}(t)$ vs time are shown in (b), (d) and (f) respectively.}
\end{center}
\end{figure*}

For one and two dimensional array, the allowed number of coupled cells are two and four respectively. So for the cells in this configuration, the delta-notch signaling activation rate, $\nu_c$ plays switching on or off of signal processing among the cells and in "on switch" condition, coupling in any cell is done via coupling of its nearest neighbour cells. Therefore in delta-notch signaling process, we have the following coupled oscillators,
\begin{eqnarray}
\frac{dM_{abc}}{dt}&=&\nu_1\Gamma_1-\nu_2\Gamma_2\nonumber\\
&&+\nu_c\Gamma_1\times\frac{1}{L}\sum_{\langle a,\alpha\rangle}\sum_{\langle b,\beta\rangle}\sum_{\langle c,\gamma\rangle}W_{\alpha\beta\gamma}\\
\frac{dY_{abc}}{dt}&=&\nu_3M-\nu_4\Gamma_3-\nu_5Y\\
\frac{dZ_{abc}}{dt}&=&\nu_5Y-\nu_6\Gamma_4\\
\frac{dW_{abc}}{dt}&=&\nu_7\Gamma_5-\nu_8\Gamma_6
\end{eqnarray}
where, $a,b,c=1,2,\dots, N$, $L=2,4,6$ for 1, 2, 3 dimensional array of cells and $\langle a,\alpha\rangle$, $\langle a,\beta\rangle$, $\langle c,\gamma\rangle$ indicate that the pairs $(a,\alpha)$, $(b,\beta)$ and $(c,\gamma)$ respectively are nearest neighbours. The mode of coupling among the cells is quite similar to mean field coupling scheme \cite{did,ros1} but the way how the coupling is incorporated among the cells in this case is localized within the nearest neighbours only. In this coupling mechanism, once the coupling is switch on with a particular value of $\nu_c$, then the cells start synchronized among themselves to exhibit coordinated behaviour. In our simulation we use fixed boundary conditions in the coupling, i.e. $W_{0bc}=0$, $W_{a0c}=0$, $W_{ab0}=0$; $W_{N+1,bc}=0$, $W_{a,N+1,c}=0$, $W_{ab,N+1}=0$. 

It has been pointed out that the identification of phase synchrony of any two identical systems can be done qualitatively by the measurement of the time evolution of instantaneous phase difference of the two systems \cite{sak,pik,ros,nan}. It is possible if one defines an instantaneous phase of an {\it arbitrary} signal $x(t)$ via Hilbert transform \cite{ros}
\begin{eqnarray}
\label{phase}
\tilde x(t)=\frac{1}{\pi}P. V. \int_{-\infty}^{+\infty}\frac{x(t)}{t-\tau}d\tau
\end{eqnarray}
where $P. V.$ denotes the Cauchy principal value. Then, one can determine an instantaneous "phase" $\phi(t)$ and an instantaneous "amplitude" $A(t)$ of the given signal through the relation, $A(t)e^{i\phi(t)}= x(t)+i\tilde x(t)$. Given two signals of two systems $x_i(t)$ and $x_j(t)$, one can therefore obtain instantaneous phases $\phi_i(t)$ and $\phi_j(t)$; phase synchronization is then the condition that $\Delta\phi_{ij}=m\phi_i-n\phi_j$ is constant, where $m$ and $n$ are integers. Of most interest are the cases $\Delta \phi_{ij}$ = 0 or $\pi$, namely the cases of in--phase or anti--phase, but other temporal arrangements may also occur. 

The phase synchronization of the two identical systems can also be identified by doing pecora-caroll type recurrence plot of the variable of one system, say $x$ with the corresponding variable $x^\prime$ of the other system simultaneously on two dimensional cartesian co-ordinate system \cite{pec}. The rate of synchronization between the two systems can be determined by the rate of concentration of the points in the plot along the diagonal. If the two systems are uncoupled, then the points on the plot will scatter away randomly from the diagonal.
\begin{figure*}
\label{fig6}
\begin{center}
\includegraphics[height=320 pt]{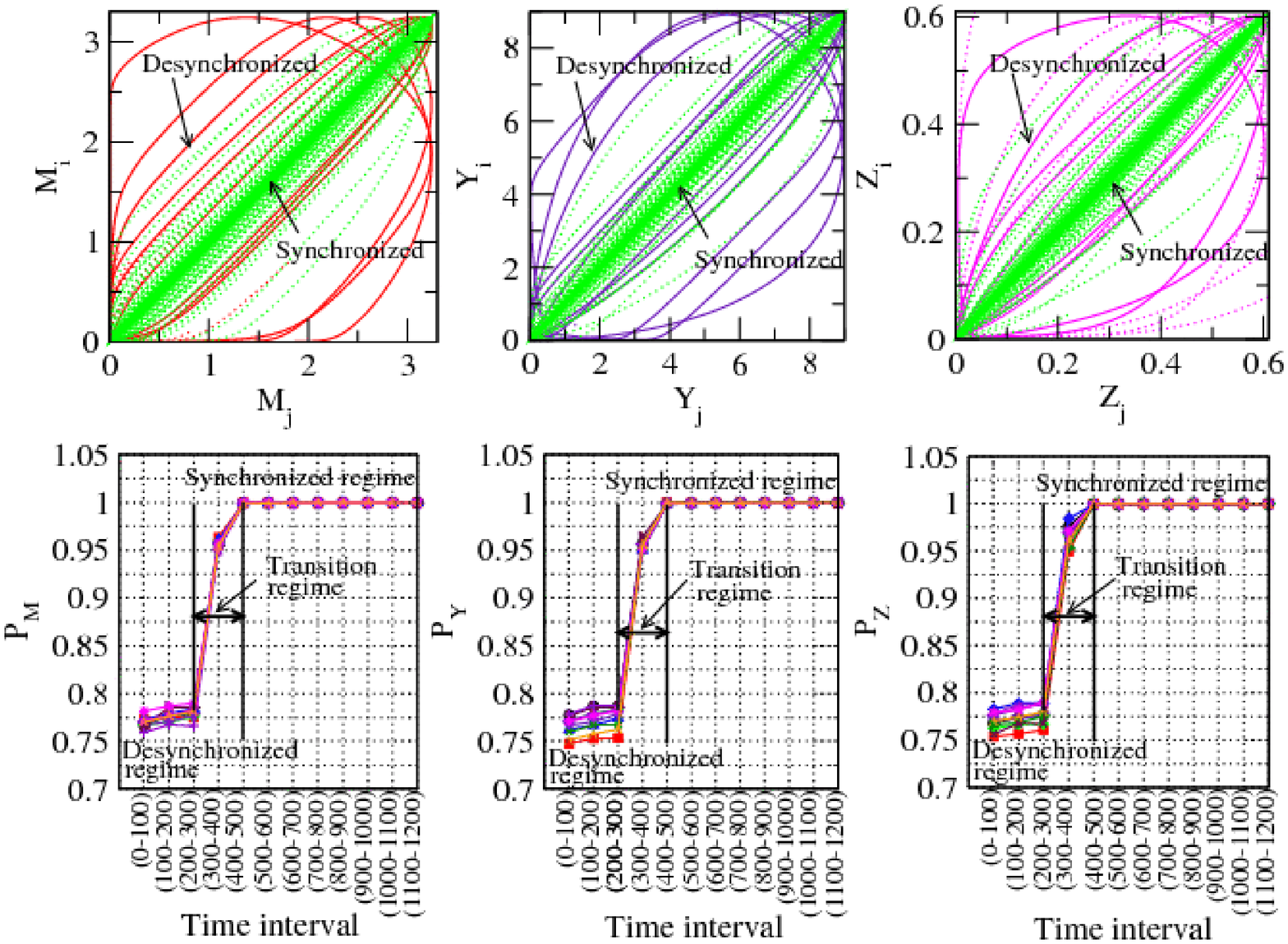}
\caption{The pecora-caroll type recurrence plots of the $M$, $Y$ and $Z$ of 10 oscillators shown in Fig. 3 are shown in the upper three panels. $i,j=1,2,...,10$, $i\ne j$. The corresponding $P$ values of the three variables for different time periods are shown in the lower three panels.}
\end{center}
\end{figure*}

The rate of phase synchrony of the two systems can be estimated quantitatively by measuring the "phase locking value" (PLV) of the two signals of the two systems \cite{lac}. The phase locking value, which is used to quantify the degree of synchrony, characterizes the stability of phase differences between the phases $\phi_i(t)$ and $\phi_j(t)$ of two signals $x_i(t)$ and $x_j(t)$ of ith and jth systems and can be defined within a time period $T$ by,
\begin{eqnarray}
\label{plv}
P(t)=\frac{1}{T}\left|\sum_{t-T}^t\Delta\phi_{ij}\right|
\end{eqnarray}
The range of PLV value is [0,1]. When the value of PLV is zero, the two systems are uncoupled, whereas if the value of PLV is one then the two systems are perfectly phase synchronized \cite{lac}.

\section{Results and discussions}

We first present one time simulation results of $M$, $Y$ and $Z$ of a single cell as a function of time for the values of $\nu_c=0.263,0.5$ and 0.7 respectively in Fig. 3 by solving differential equations 1,2,3 and 4 by standard 4th order Runge Kutta numerical integration method \cite{pre}. In the 3D plot, the dynamics shows three different limit cycle oscillations for these three different $\nu_c$ values. In Fig. 4, the corresponding dynamics of the three variables for two coupling cells are shown in panels (a), (c) and (e) indicating phase synchrony irrespective of variation in $\nu_c$. The claim is supported by the phase plots i.e. $\Delta\phi_{12}$ vs time in Fig. 4 (b), (d) and (f), along with pecora-caroll type recurrence plots in the panels (g), (h) and (i). It is noticed from the plot that the amplitudes of oscillations of all variables are increased as $\nu_c$ is increased however the frequencies of oscillations are decreased correspondingly. The phase locking value for whole time period (0-1500) minutes using equation (\ref{plv}) is found to be 0.97$\pm$0.015 indicating fairly well synchronized. There are signatures of fluctuations in the beginning of each plot due to the cells evolve in different initial conditions and takes some time to synchronize. This means that once the coupling is switch on via $\nu_c$ among the cells, the variation in $\nu_c$ will hardly effect in the synchrony rate of the cells but will occur at different amplitudes.

Next we take $(10\times 10\times 10)$ such cells arranged in three dimensional array. We take $\nu_c=0.263$ and apply fixed boundary conditions and simulate differential equations 5, 6, 7, 8 and the results of $M$, $Y$ and $Z$ as a function of time for 10 cells are shown in Fig. 5. The coupling variable for each cell is $W$ via mean field like coupling mechanism restricting to nearest neighbours only and the coupling is switch on at time 300 minutes. In the plots of Fig. 5 (a), (c) and (e), three regimes i.e. desynchronized, transition and synchronized regimes are seen clearly. The phase transition like behaviour is supported by the phase diagrams ($\Delta\phi_{ij}$ vs time) of pairs of cells in figures (b), (d) and (f) where desynchronized regime is indicated by random evolution of $\Delta\phi_{ij}$ with time, transition regime is identified by weak fluctuation of $\Delta\phi_{ij}$ around a constant value and synchronized regime is indicated by constant $\Delta\phi_{ij}$ with time. Again this claim of phase transition is verified qualitatively by pecora-caroll recurrence plots in the upper three panels of Fig. 6. In these plots if the cells are desynchronized, the points in the plots will spread away from the diagonal, however in the transition regime the points start concentrating towards the diagonal. When the points lie along the diagonal the cells are synchronized as shown in the plots.

Fixing the configuration of cells, we calculated $\Delta\phi_{ij}$ of pair of far distant cells (i.e. i=1, j=10) as a function of time, which are already included in the respective diagrams, and found the same pattern of phase transition. This means that once the coupling is switch on, all the cells process information almost instantaneously even at far distant cells also in the configuration. Hence long range information transfer or relay is done instantaneously when the cells are coupled.

The quantitative identification of this phase transition can be explained more clearly by the measurement of phase locking values ($P_M$, $P_Y$ and $P_Z$) of the variables $M$, $Y$ and $Z$ for different time intervals by using the equation (\ref{plv}). The results are shown in the lower three panels of Fig. 6. In these plots, it is seen distinctly the demarcation of the three regimes, desynchronized, transition and synchronized regimes. In the synchronized regime, the values of $P_M$, $P_Y$ and $P_Z$ are 0.994$\pm$0.002 starting from time interval (400-500) minutes to (1100-1200) minutes. In the time intervals before switching the coupling, the values of $P_M$, $P_Y$ and $P_Z$ are in 0.77$\pm$0.05 indicating desynchronized regime. But in the time intervals 300 to 500 minutes, the values of $P_M$, $P_Y$ and $P_Z$ monotonically increased indicating transition regime.

\section{Conclusion}

The phase synchrony shown by temporal oscillations of the variables indicates that once the coupling is switched on with the mean field like coupling mechanism with a value of activation rate $\nu_c$ above some critical value to exhibit synchrony, then the increase of $\nu_c$ does not affect much in the rate of phase synchrony. However it affects in the magnitudes of amplitudes and frequencies of oscillations of the dynamical variables of the cells. It is also to be noted that if the variables evolve with time from different initial conditions, after coupling is switched on, it takes some time to get synchronized. 

Since the cells coupled only when they are very close or in contact and the delta protein cannot diffuse from one cell to another, the mean field like coupling scheme is the most reasonable way of signal processing in this situation. We take three dimensional array of cells to see the behaviour of phase synchrony in more real situation. Our results interestingly show phase transition like behaviour by identifying desynchronized, transition and synchronized regimes. Relay information transfer among the far cells is done among the cells themselves once the coupling is introduced in a group of cells. To investigate it in more real situation one has to incorporate noise fluctuations arising from internal molecular events and external environmental fluctuations.

\section{Acknowledgments}
This work is financially supported by Department of Science and Technology (DST) and carried out in Centre for Interdesciplinary Research in Basic Sciences, Jamia Millia Islamia,New Delhi,India.

\end{document}